\documentclass{article}

\usepackage{amssymb}
\usepackage{graphicx}
\usepackage{bm}
\usepackage{mathtools}
\usepackage{color}
\usepackage{cases}
\usepackage{amsthm}

\title{Neural vortex method: from finite Lagrangian particles to infinite dimensional Eulerian dynamics}
\author{
  Shiying Xiong \\
  Zhejiang University and Dartmouth College, shiying.xiong@zju.edu.cn\\
  Xingzhe He, Yunjin Tong, Yitong Deng, and Bo Zhu\\
  Dartmouth College\\
}

\begin{document}

\maketitle

\begin{abstract}
In fluid analysis, there has been a long-standing problem: lacking a rigorous mathematical tool to map from a continuous flow field to finite discrete particles, hurdling the Lagrangian particles from inheriting the high resolution of a large-scale Eulerian solver.
To tackle this challenge, we propose a novel learning-based framework, the neural vortex method (NVM). NVM builds a neural-network description of the Lagrangian vortex structures and their interaction dynamics to reconstruct the high-resolution Eulerian flow field in a physically-precise manner.
The key components of our infrastructure consist of two networks: a vortex detection network to identify the Lagrangian vortices from a grid-based velocity field and a vortex dynamics network to learn the underlying governing interactions of these finite structures.
By embedding these two networks with a vorticity-to-velocity Poisson solver and training its parameters using the fluid data obtained from grid-based numerical simulation, we can predict the accurate fluid dynamics on a precision level that was infeasible for all the previous conventional vortex methods.
We demonstrate the efficacy of our method in generating highly accurate prediction results with low computational cost by predicting the evolution of the leapfrogging vortex rings system, the turbulence system, and the systems governed by Navier--Stokes (NS) equations with different external forces. We compare the prediction results made by NVM and the Lagrangian vortex method (LVM) for solving the NS equation in the periodic box and find that the relative error of the predicted velocity using NVM is more than 10 times lower than that of the LVM. Moreover, our method only requires data collected from a very short training window, more than 100 times smaller than the prediction period, which potentially facilitates data acquisition in real systems.
\end{abstract}

\section{Introduction}\label{sec:into}
Due to the high number of degrees of freedom in the motion space, the complex nonlinear coupling between particles, and the susceptibility to numerical dissipation, reproducing the behavior of fluids with such features in a physically-accurate manner presents a challenge.
Two general approaches are developed to solve the fluid equations, one is grid-based, or Eulerian, and the other is particle-based, or Lagrangian. Conventionally, many pieces of research are done on grid-based methods mainly to achieve high resolution, for example on channel flows \cite{KimMoinMoser1987,ZhouAdrianBalachandarEtAl1999}, boundary layers \cite{lu2012numerical,wu2017transitional,Sayadi2013}, isotropic turbulence \cite{ishihara2016energy,JimenezWraySaffmanEtAl1993,JimenezWray1998}, and pipe flows \cite{Wu2015}. However, the main drawback of grid-based methods is their high computational cost that is unaffordable under many conditions \cite{Xiong2022Clebsch}. On the contrary, the particle-based approach is efficient in computation but suffers from inaccuracy in results \cite{LindsayKrasny2001,losasso2008}.

Lagrangian vortex methods (LVMs) are well-known particle-based methods which use vortices as computational elements, mimicking the physical structures in turbulence. In LVMs, the Navier--Stokes (NS) equations are formulated in terms of vorticity in contrast to the conventional formulations with velocity and pressure and are solved using a Lagrangian approach instead of the Eulerian formulation \cite{leonard1980vortex,Cottet2000,anderson1985vortex,Selle2005}. Discretizing the NS equations with finite particles results in a set of ordinary differential equations (ODEs) for the particle strengths and positions \cite{xiong2021incompressible}. The Lagrangian method has some advantages, such as its automatic adaptivity of the computational elements, the numerical conservation of physically conserved quantities, the ability to simulate phenomena covering many orders of magnitude, and the rigorous treatment of boundary conditions at infinity, in the numerical simulation of fluids in unsteady flows where the vortex structures play a leading role. For many typical flows, such as wall-bounded flows \cite{Zhao2018,di2021direct}, wake flows \cite{tong2020characterizing,mimeau2021comparison,karimi2021numerical}, and jet flows \cite{you2017effects}, their vorticity distributions are usually concentrated in certain regions, which enables us to distribute the density of the vortex elements precisely. Besides, with Kelvin's circulation theorem directly embedded into the Lagrangian dynamics of the vortex elements \cite{Hao2019}, the conservativeness during the numerical calculation can be reasonably guaranteed. Finally, we can obtain the incompressible velocity field through the discrete Biot--Savart (BS) law using vortex elements, showing a weak coupling between evolutionary quantity and the solenoidal condition of velocity, which can be beneficial to maintain the computational stability in fluid simulations in the face of physical discontinuities.  \cite{saye2016interfacial,Yang2021clebsch}.

However, the implementation of the LVM faces a major challenge which is to model the right-hand sides (RHSs) of the set of ordinary differential equations based on the NS equations. Firstly, the assumption that the vortices are point-like largely limits the use of the continuous BS law. Second, the drift velocity due to the external force cannot be obtained using the LVM without knowing the function of the external force. Even given the function, the LVM still fails to capture the drift velocity accurately in most cases \cite{Hao2019}.
Finally, when two particles are close enough, the singularity of the discrete BS law leads to a significant numerical error.
The above problems make the LVM inaccurate and inapplicable in solving the underlying fluid dynamics under many situations \cite{Cottet2000}.

While the development of traditional numerical methods facing obstacles, fortunately in the last decade, with the drastic advancement in computational power and data availability, we are presented with the new hope of approaching these previously elusive systems from a new angle: a data-driven one powered by machine learning\cite{raissi2018multistep,gonzalez1998identification,duraisamy2019turbulence,xie2018tempogan,chu2017data}. Since brute-force machine learning with conventional toolkits such as deep neural networks typically suffers from the high dimensionality of input-output spaces, expensive cost of data, the tendency to yield physically implausible results, and the inability to robustly handle extrapolation, recent research interests have been focused on embedding physical priors in learning algorithms so that the networks approach the data not as wide-eyed infants but as physicists familiar with the fundamental rules of how our world operates \cite{anderson1996comparison,crutchfield1987equations, daniels2015automated,wang2017physics,hammond2022machine,xu2022towards}. Recently, many pieces of research are developed to efficiently learn fluid dynamics by incorporating physical priors into the learning framework, e.g., encoding the
NS equations \cite{Raissi2018}, embedding the notion of an incompressible fluid \cite{mohan2020}, and identifying a mapping between the dynamics of wave-based physical phenomena and the computation in a recurrent neural network \cite{Hughes2019}.

In the realm of learning complex fluid dynamics, various works have been proposed along this line of thinking, seeking to engrave the structure of partial differential equations (PDEs) into the network architectures  \cite{yang2019predictive,raissi2020hidden,belbute2020combining,lye2020deep,white2019neural,mohan2020embedding}. Ideally, the PDEs should learn to evolve the flow field without consulting the particular solver to obtain initial-condition invariance. Still, due to the high dimensionality and the lack of supervision, the machine-learning community has not solved such a task to date.

In this work, we propose a novel approach, the neural vortex method (NVM), to embed prior physical knowledge to elegantly achieve invariance to initial conditions while at the same time being efficient in data usage, easy to implement, fast to compute, adaptable to arbitrarily high precision, and suitable for handling complex, unsteady flows.
Our method is inspired by the vortex method in fluid dynamics, which discretizes the NS equations with a set of Lagrangian particles ----- vortices, based on Helmholtz's theorems which states that the behavior of the fluid can be described by several vortex elements flowing with the fluid \cite{barba2004vortex}. We remark that accurate simulation of complex vortex dynamics requires a large number of Lagrangian vortex particles, considering the spatially continuous distribution of vortex structures. However, significant theoretical and numerical errors may occur if we only advect the fluid with a small number of vortex particles without considering vortex deformation, external forces, and viscous dissipation. We aim to establish a method to reconstruct continuous vortex dynamics with a small number of vortex particles. We model the position and dynamics of the vortex particles in a data-driven manner, and errors can be significantly reduced compared with a direct formula-based model.

In the same spirit as the LVM, our method learns to describe a complex fluid dynamics system by first learning to associate a small number of Lagrangian vortex particles to the observation and simulate the forward dynamics of these vortices instead. This approach can be viewed as identifying a compact, physics-based latent space where simulation can be performed efficiently.
To view it another way, instead of tailoring our networks to solve the governing PDEs, we design our networks to describe the underlying behavior of our system that such equations are proven to imply.
We show the superiority of our vortex-based approach to the previous approaches with its ability to generalize to different initial conditions, adapt to arbitrary precision, learn with small data sizes, and simulate complex turbulent flows.

Our method brings a new approach to identifying fluid systems exhibiting complex vortical motions. Still, its idea can be enlightening to other fields as well.
We map continuous flow fields to low-dimensional Lagrangian vortex dynamics to study fluid evolution over time, which can also be generalized to common physical systems. In general, we can map the physical space described by PDE to a low-dimensional discrete space through a neural network to predict the time evolution of the transformed system and recover to the original physical space through neural networks or physical prior. We remark that to better preserve the critical structures, the intrinsic geometric symmetry should be taken into account when transforming the physical system into a different space.
As incorporating physical priors is an imminent and promising trend, with this schematically novel approach, our work can potentially open up a new horizon for future endeavors.

The structure of this paper is as follows. In section \ref{sec:LVM}, we will first introduce the NS equations and LVM that serve as the mathematical foundation of our methodology. Then, we give an overall of our method in \ref{sec:nvm}. The following section \ref{sec:set} describes the details of our dataset generation and training settings. In section \ref{sec:result}, we show the numerical results, which compare our methodology with
the Lagrangian method.
Lastly, conclusions are drawn in section \ref{sec:concl}, with an emphasis on the significance of our method.

\section{NS equations and LVM}\label{sec:LVM}

\subsection{NS equations}

Given a fluid velocity field $\bm u (\bm x,t)$ with an incompressible constraint, its underlying dynamics can be described by the NS equations
\begin{equation}
  \begin{dcases}
  \frac{D \bm{u}}{D t}=-\frac{1}{\rho}\bm{\nabla} p+\nu\nabla^2 \bm u +\bm f,\\
  \bm{\nabla}\cdot \bm{u}=0,
  \end{dcases}
  \label{eq:uNS}
\end{equation}
where $t$ denotes the time, $D/D t = \partial / \partial t + \bm u \cdot \bm \nabla $ is the material derivative, $p$ is the pressure, $\nu$ is the kinematic viscosity, $\rho$ is the density, and $\bm f$ is the body accelerations (per unit mass) acting on the continuum, for example, gravity, inertial accelerations, electric field acceleration, and so on.

The alternative form of the NS equations could be obtained by defining the vorticity field $\bm{\omega} = \bm{\nabla \times u}$, which leads to the following vorticity dynamical equation
\begin{equation}
\begin{dcases}
  \frac{D \bm{\omega}}{D t}=(\bm{\omega}\cdot \bm{\nabla})\bm{u} + \nu\bm{\nabla}^{2}\bm{\omega} + \bm{\nabla} \times \bm f,\\
  \nabla^2 \bm \Psi = -\bm \omega,~~\bm u = \bm \nabla \times \bm \Psi,
\end{dcases}
  \label{eq:wNS}
\end{equation}
where $\bm \Psi$ is a vector potential whose curl is the velocity field.
Although this form does not seem to bring any simplification, the key illumination of doing this transformation stems the Helmholtz's theorems,\cite{Helmholtz1858} which states that the dynamics of the vorticity field can be described by vortex surfaces/lines, which are Lagrangian surfaces/lines flowing with the velocity field in inviscid flows \cite{Yang2010b,Xiong2017}.

\subsection{LVM}

The LVM discretizes the vorticity dynamical equation \eqref{eq:wNS} with $N$ particles resulting in a set of ODEs for the particle strengths $\bm \Gamma = \{\bm \Gamma_i|i=1,\cdots,N\}$ and the particle positions $\bm X =\{\bm X_i|i=1,\cdots,N\}$ as
\begin{equation}
  \begin{dcases}
  \frac{\textrm{d} \bm \Gamma_i}{\textrm{d} t}= \bm \gamma_i,\\
  \frac{\textrm{d} \bm X_i}{\textrm{d} t}= \bm u_i + \bm v_i.
  \end{dcases}
  \label{eq:vortex}
\end{equation}
Here, the particle strength $\bm \Gamma_i$ is the integral of $\bm \omega$ over the $i^{\textrm{th}}$ computational element,
$\bm u_i$ is the induced velocity calculated by BS law
\begin{equation}
    \bm u_i = \frac{1}{2(n_d-1)\pi}\sum_{j\neq i}^{N} \frac{\bm \Gamma_j\times (\bm X_i - \bm X_j)}{|\bm X_i - \bm X_j|^{n_d}+ \mathcal{R}^{n_d}},
    \label{eq:BS}
\end{equation}
where $n_d$ is the dimension of the flow field.
In addition, $\bm \gamma_i$ and $\bm v_i$ are the change rate of the particle strength and the drift velocity \cite{Hao2019}, respectively. 
To avoid singularities in the BS law, we introduce the numerical regularization parameter $\mathcal{R}$ in the LVM as $\mathcal{R}=0.1$. The effect of the regularization parameter on the dynamics of the flow evolution of the simulated vortex particles is rather small because of the large spacing between the vortex particles.

In a two-dimensional ideal fluid flow, i.e., a strictly inviscid barotropic flow with conservative body forces, the movements of Lagrangian particles with conserved vorticity strength are determined by the velocity field they create, thus allowing us to advance the simulation temporally \cite{Cottet2000}. However, in the real three-dimensional flow, under the action of vortex stretching, vortex distortion, viscous dissipation, external forces, etc., the Lagrangian advection of vortex particles and their strength need to be corrected by $\gamma_i$ and $\bm v_i$ in \eqref{eq:vortex}.

We remark that the NS equations can be accurately modeled by the Lagrangian vortex method with a large number of computational elements and a reasonable discrete distribution. Optimizing the discrete distribution and dynamics modeling by neural networks deserves further research.

\section{NVM} \label{sec:nvm}

\subsection{Methodology}\label{subsec:meth}

To quantify the fluid dynamics accurately and efficiently, we propose a novel framework, the NVM, that extracts information from the Eulerian specification of the flow field (or the images of flow visualizations) and translates it into knowledge about the underlying fluid field through physics-informed neural networks.
To build a fully automated toolchain that can extract a high-resolution Eulerian flow field from the Lagrangian inductive priors, we embed these two networks with a vorticity-to-velocity Poisson solver, as shown in Figure \ref{fig:overview}.
The reason behind this structure is that learning directly from high-dimensional observations, such as images, is unable to be achieved using traditional methods since extracting the velocity and pressure fields directly from the images is challenging.

\begin{figure}
  \centering
  \includegraphics[width=1.\textwidth]{overview.eps}\\
  \caption{Schematic diagram of the NVM. Our system is constituted of two networks, the detection network and the dynamics network, which are embedded with a vorticity-to-velocity Poisson solver.}
  \label{fig:overview}
\end{figure}

We construct a vortex detection network in section \ref{subsec:detnet} to identify the positions and the vorticity of Lagrangian vortices from a grid-based velocity field, which from a mathematical perspective connects \eqref{eq:uNS} with \eqref{eq:vortex}. In this way, we simplify the vorticity field into a field only constituted of the identified vortices. Given the detected vortices, we then use a vortex dynamics network in section \ref{subsec:dynet} to learn the underlying governing dynamics of these finite structures. Dynamics networks accurately model the RHSs of \eqref{eq:vortex} under various conditions, resolving the long-standing problem in LVM.

 We train the parameters using high-fidelity data from high-resolution direct numerical simulation. The model is trained only with information collected from the interaction of 2 to 6 vortices. The trained model can be applied to any arbitrary vorticity field with any number of vortices.
The training process of NVM consists of 2 major steps: detection network training and dynamics network training. we utilize data collected from randomly generated vortices and the corresponding vorticity fields to train the detection network.
Direct numerical simulation (DNS) is used to calculate the data of the evolutionary vorticity fields. We use the well-trained detection network to identify vortices' positions and vorticity strength from the initial and the evolved vorticity fields. The identified vortices are then used to train the dynamics network.

\subsection{Detection network}
\label{subsec:detnet}

The input of the detection network is a vorticity field of size $200\times200\times1$. As shown in Figure \ref{fig:detection}, we first feed the vorticity field into a small one-stage detection network and get the feature map of size $25\times25\times512$ (we downsampled 3 times). The detection network consists of a Conv2d-BatchNorm-ReLU combo and a 6-layer-structured ResBlock chain whose size can be adjusted dynamically to the complexity of the problem. The primary reason for downsampling is to avoid extremely unbalanced data and multiple predictions for the same vortex. We then forward the feature map to 2 branches. In the first branch, we conduct a $1\times1$ convolution to generate a probability score $\hat{p}$ of the possibility that there exists a vortex. If $\hat{p} > 0.5$, we believe there exists a vortex within the corresponding cells of the original $200\times200\times1$ vorticity field. In the second branch, we predict the relative position to the left-up corner of the cell of the feature map if the cell contains a vortex. Afterward, we set a bounding box of $10\times10$ around these
predicted vortices and use the weighted average of the positions of the cells of the original vorticity field to find the exact position of the vortex. Finally, the vortex particle strength is calculated as the sum of the value of the cells in the bounding box normalized by the cell area.

\begin{figure}
  \centering
  \includegraphics[width=1.\textwidth]{detection.eps}
  \caption{The architecture of the detection network. It takes the vorticity field as input and outputs the position and vortex particle strength for each vortex detected. The \textit{Conv} means the Conv2d-BatchNorm-ReLU combo, and the \textit{ResBlock} is the same as in \cite{he2016resnet}. In each \textit{ResBlock}, we use stride 2 to downsample the feature map. The Resblock chain is six-layer structured.} The number in the parenthesis is the output dimension.
  \label{fig:detection}
\end{figure}

In the training process, we penalize the wrong position detection only if the cell containing a vortex in the ground truth given by DNS is not detected. This idea is similar to the real-time object detection in \cite{redmon2016yolo}. We do not use the weighted average method to find the position in the training to ensure the detection network can produce detection results as accurately as possible. We use the focal loss \cite{lin2017focal} to further relieve the unbalanced classification problem.

We mainly use the detection network to generate training data for the dynamics network because we want to use the high-resolution data generated by the method mentioned in section \ref{subsec:data_gen} instead of by the approximate particle method (BS law). Moreover, there are many situations where BS law is inapplicable, as discussed previously in section \ref{sec:into}. The detection network enables us to find the positions of the vortices accurately regardless of the situation.

\begin{figure}
  \centering
  \includegraphics[width=1.\textwidth]{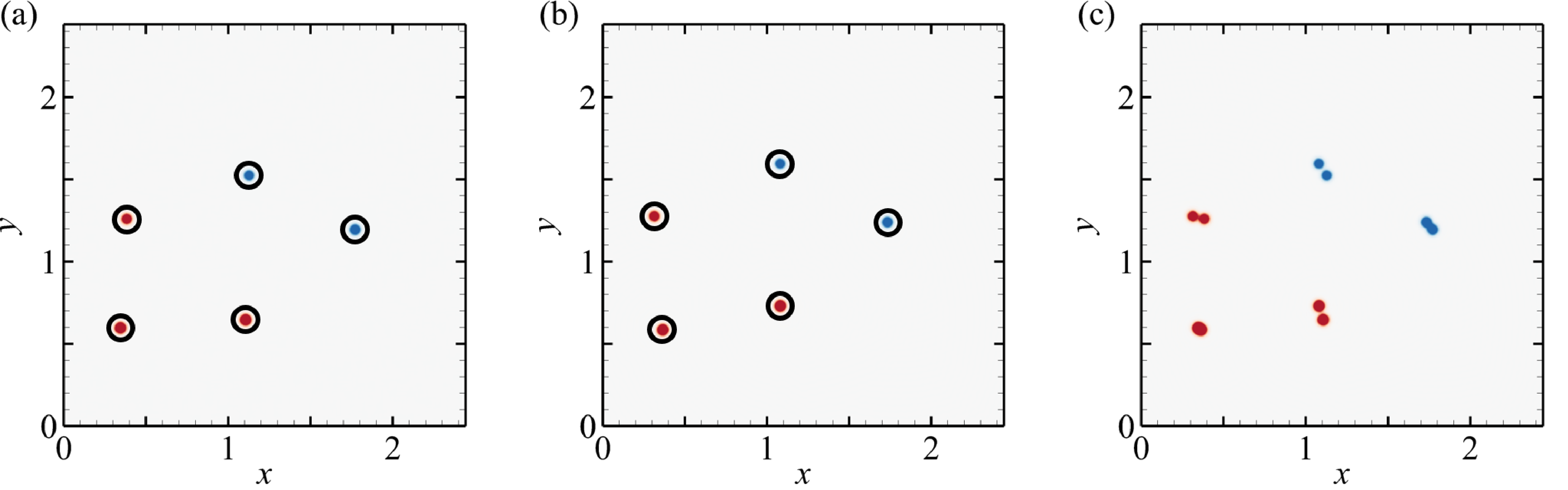}\\
  \caption{A example of vorticity contour at (a) $t=0$, (b) $t=0.2$, and (c) superposition of $t=0$ and $t=0.2$. The black circles indicate the location recognized by the detection network. The evolution from (a) to (b) is calculated by DNS.}
  \label{fig:detection123}
\end{figure}

The detection network is responsible for providing necessary information to the dynamics network. After the training, we use the well-trained detection network to detect the vortices in the initial vorticity fields and the evolved vorticity field, both generated by the method in section \ref{subsec:data_gen}. We then apply the nearest-neighbor method to pair the vortices detected in these two fields. Figure \ref{fig:detection123} shows the case of two fields at $t=0$ and $t=0.2$. The idea of nearest-neighbor pairing can be perceived from Figure \ref{fig:detection123} (c). The sample, or these two fields, is dropped if different numbers of vortices are detected in the initial and evolved fields or if a large difference exists in the vorticity of paired vortices. The successfully detected vortices in the initial and evolved vorticity fields are passed together into the dynamics network for its training.

We remark that our method does not venture into the detection of three-dimensional vortex structure. It is more challenging to localize vortices in complex three-dimensional structures exhibiting tubular or sheet-like morphology by neural networks. A 3-D vortex identification model requires a large number of vortex particles, to reduce its computational cost, a promising direction is to treat vortex elements as line segments with length and direction as in \cite{xiong2021incompressible}.

\subsection{Dynamics network}
\label{subsec:dynet}

\begin{figure}
  \centering
  \includegraphics[width=1.\textwidth]{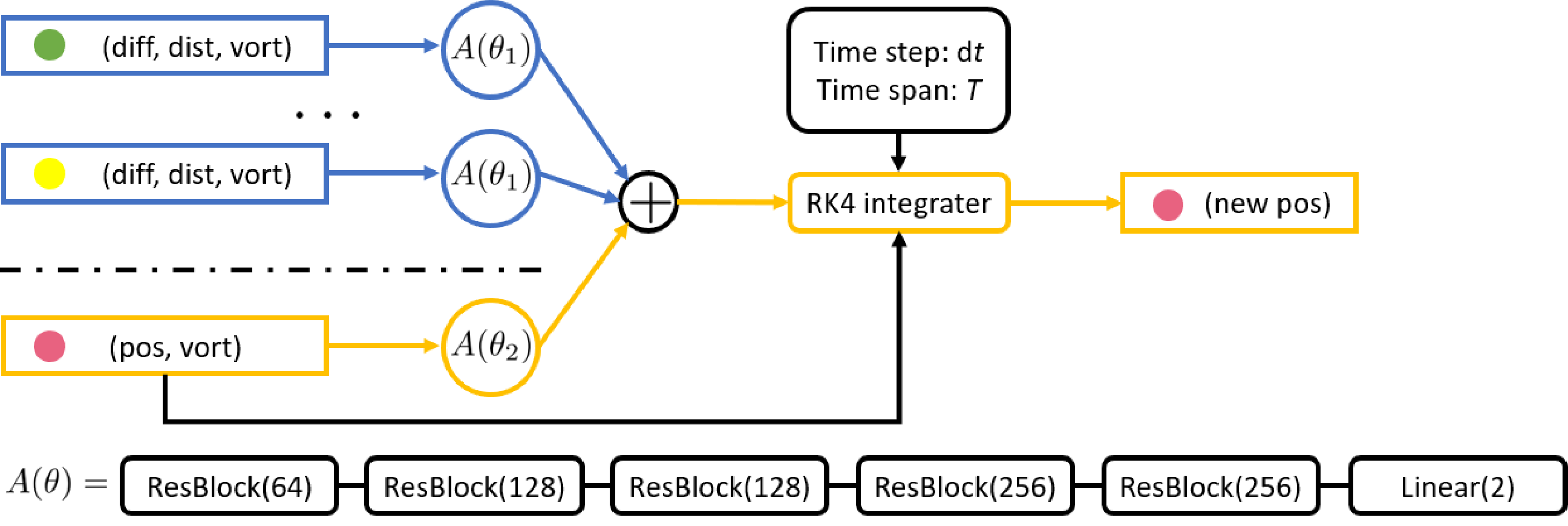}
  \caption{The architecture of the dynamics network. It takes the particle's attribution as input and outputs each vortex's position. The \textit{ResBlock} has the same architecture as in \cite{he2016resnet} with the convolution layers replaced by linear layers. The number in the parenthesis is the output dimension.}
  \label{fig:dynamics}
\end{figure}

To learn the underlying dynamics of the vortices, we build a graph neural network similar to \cite{battaglia2016interaction}. We predict the velocity of one vortex due to influences exerted by the other vortices and the external force. Then we use the fourth-order Runge--Kutta integrator to calculate the position in the next timestamp. As shown in Figure \ref{fig:dynamics}, for each vortex, we use a neural network $A(\theta_1)$ to predict the influences exerted by the other vortices and add them up. Specifically, for each $i$th vortex, we consider the vortex $j (j\neq i)$. The difference of their positions can be calculated by $\textrm{diff}_{ij}=\textrm{pos}_i-\textrm{pos}_j$, and their L2 distance is $\textrm{dist}_{ij}=\|\textrm{diff}_{ij}\|_2$. The input of the $A(\theta_1)$ is the vector $(\textrm{diff}_{ij}, \textrm{dist}_{ij}, \textrm{vort}_j)$ of length 4. Here, $\textrm{pos}$ and $\textrm{vort}$ are detected by the detection network.
The output of $A(\theta_1)$ is a vector with the same dimension of the flow field, characterizing the induced velocity of the $j$th vortex to the $i$th vortex. In this way, we can calculate the induced velocity of each vortex $j$ ($j\neq i$) on the vortex $i$. We sum up all the induced velocities on the vortex $i$ and treat the result as the induced velocity exerted by the other vortices.

In addition, we use another neural network $A(\theta_2)$, to predict the influence caused by the external force, which is determined by the local vorticity and the position of the vortex. The input of $A(\theta_2)$ is a vector of length 3. The output is the influence exerted by the environment on the vortex $i$, i.e., the induced velocity of the external force to $i$th vortex.

The reason we separate the induced velocity into two parts, i.e., $A(\theta_1)$ and $A(\theta_2)$, is as follows. On the one hand, the induced velocities between vortex particles are global, and exhibit a certain symmetry, i.e., the vortex particles interact with each other following the same law. In contrast, the influence of external forces on vortex particles is usually local and direct; thus, we do not need to consider the interaction between particles. The effect of the vortex stretching term in three-dimensional vortex flows or diffusion term in viscous flows is also local and should be included in network $A(\theta_2)$. Note that both the outputs of $A(\theta_1)$ and $A(\theta_2)$ are a vector with the same dimension of the flow field. Thus, we can add the two kinds of influence together, whose result is defined as the velocity of the vortex $i$. We feed the velocity into the fourth-order Runge--Kutta integrator to obtain the predicted position of vortex $i$.

In addition, in predicting the evolution of the flow field, NVM replaces the discrete BS method with a dynamics network composed of ResBlocks. We chose a 5-layer ResBloks to improve the expressiveness of the dynamics network so that we can learn dynamics of different complexity on the same network. Since the dynamics network with 5-layer ResBloks is more complex than the discrete BS method, the computational cost of NVM is higher than that of the Lagrangian vortex method. We remark that although the computational cost of ResBlocks itself is relatively large in NVM, the number of vortex particles needed to predict the evolution of the flow field using NVM is much smaller. Therefore, the overall computational cost of NVM can be greatly reduced.

\section{Dataset generation and training settings}\label{sec:set}

\subsection{Dataset generation and tainning settings}\label{subsec:data_gen}
We randomly sample 2 to 6 vortices and create the initial vorticity field through convolution with a Gaussian kernel $\sim\mathcal{N}(0,0.01)$. This process is repeated 2000 times to generate $N_s = 2000$ samples. DNS is performed to solve \eqref{eq:uNS} in the periodic box using a standard pseudo-spectral method \cite{Rogallo1981}.
Aliasing errors are removed using the two-thirds truncation method with the maximum wavenumber $k_{\max}\approx N/3$.
The Fourier coefficients of the velocity are advanced in time using a second-order Adams--Bashforth method. The time step is chosen to ensure that the Courant--Friedrichs--Lewy number is less than $0.5$ for numerical stability and accuracy. To obtain accurate DNS data samples, we set the grid size as $N=1024$. Regarding the kinematic viscosity, we set $\nu=0$ for the cases in section \ref{sec:threeforce} and $\nu=0.001$ in the other cases.
The pseudo-spectral method used in this DNS is similar to that described in
\cite{Xiong2017,Xiong2019,Xiong2020}.

We use $N_{train}= 0.9 N_s = 1600$ samples with the time span $T_{train}$ for the training of the dynamics network. The DNS dataset is generated with random initial conditions independent of the predicted vortex evolution. The time step of vortex evolution is set as $\textrm{d}t$. For the leapfrog example, we set the parameters as $T_{train}=1$ and $\textrm{d}t=0.001$.
For the turbulent flow example, we set the parameters as $T_{train}=0.001$ and $\textrm{d}t=0.001$.
For other examples, the parameters are set as $T_{train}=0.2$ and $\textrm{d}t=0.1$.
In general, the parameters are chosen within a wide range, indicating the robustness of the network.
We use the trained network to predict the vortex dynamics at time $T_{predict}$. We show that the prediction time span $T_{prediction}$ is larger than the training time span $T_{train}$ in the results section, in some cases up to tens of times of $T_{train}$.

For both the detection network and the dynamics network, we use Adam optimizer \cite{kingma2014adam} with a learning rate of 1e-3. The learning rate decays every 20 epochs by a multiplicative factor of 0.8.
For the detection network, we use a batch size of 32 and train it for 350 epochs. We use the cross entropy as the classification loss and L1 loss for position prediction. To relieve the unbalanced data problem in the detection network, we implement Focal loss \cite{lin2017focal} with $\alpha=0.4$ and $\gamma=2$. It takes 15 minutes to converge on a single Nvidia RTX 2080Ti GPU.
For the dynamics network, we use a batch size of 64 and train it for 500 epochs. We use L1 loss for position prediction. It takes 25 minutes to converge on a single Nvidia RTX 2080Ti GPU.

\section{Results}\label{sec:result}

We demonstrate the efficacy of our method in generating highly accurate prediction results, with low computational cost, for the leapfrogging vortex rings system, the turbulence system, and the systems governed by NS equations with different external forces that are challenging to model for the LVM.

\subsection{Comparison between NVM and LVM}

\begin{figure}
  \centering
  \includegraphics[width=1.0\textwidth]{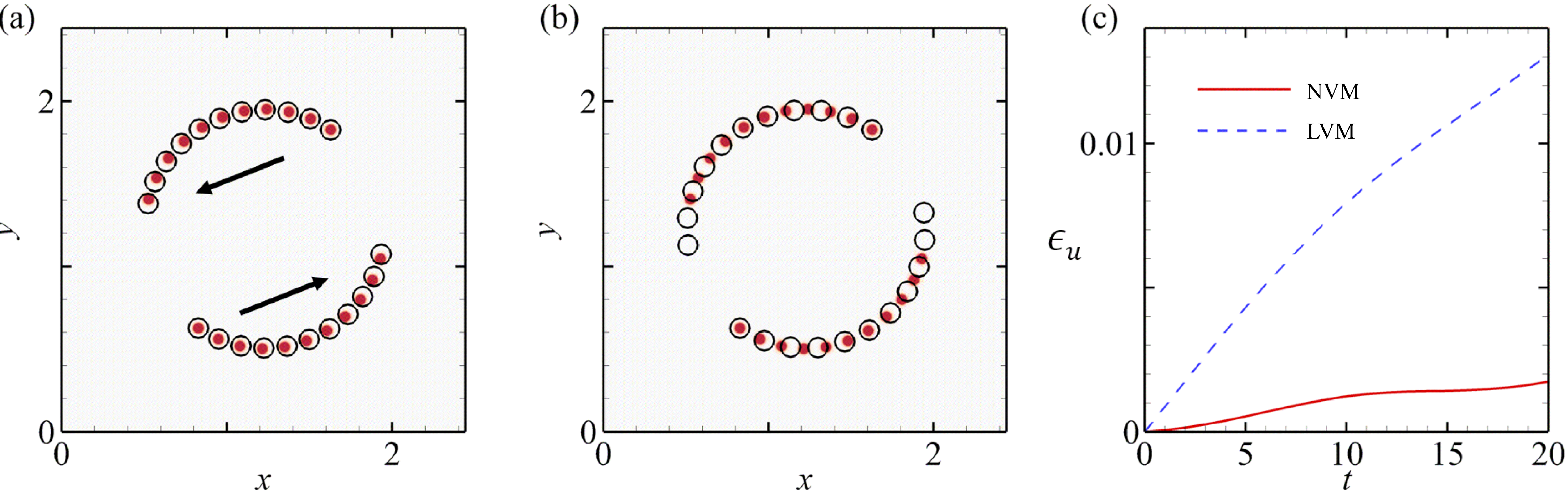}\\
  \caption{Comparison of NVM and LVM for solving NS equations in the periodic box. (a)  NVM, (b) LVM, and (c) The relative error of velocity in flow simulation. The red dots indicate the positions of 2 vortices at different time steps generated by DNS. The black circles in (a) and (b) are the prediction and simulation results of the NVM and LVMs, respectively. The black arrows indicate the directions of the motions of the 2 vortices. }
  \label{fig:uError}
\end{figure}

To demonstrate that NVM is a better approach to capturing fluid dynamics than the traditional LVM, we compare the prediction results of the NVM and the LVM for solving NS equations in the periodic box.
In the prediction, we initialize two vortex particles at $\bm X_1 = (\pi - 0.4,\pi -0.6)$ and $\bm X_2 = (\pi + 0.4,\pi +0.6)$, where the corresponding particle strength are $\Gamma_1=0.75$ and  $\Gamma_2 =0.75$.
We plot the results using the NVM and LVM and the relative error of velocity in the simulation in Figure \ref{fig:uError} (a), (b), and (c), respectively. Here, the relative error of velocity is defined as
\begin{equation}
\epsilon_u = \frac{\|\bm u_{predict}-\bm  u_{true}\|_{L^2}}{\|\bm  u_{true}\|_{L^2}},
\end{equation}
where $\bm u_{predict}$ denotes the predicted or simulated solution and $\bm  u_{true}$ denote the ground truth solution.
It is quite obvious that in Figure \ref{fig:uError} (a), the predictions made by NVM match the positions of vortices generated by DNS almost perfectly, while the predictions made by BS law in Figures \ref{fig:uError} (b) contain a large error. The divergence of the relative error of velocity is shown in Figure \ref{fig:uError} (c), which shows that the NVM outperforms traditional methods by increasing amounts as the predicting period becomes longer.

\subsection{Leapfrog vortex}
\begin{figure}
  \centering
  \includegraphics[width=1.0\textwidth]{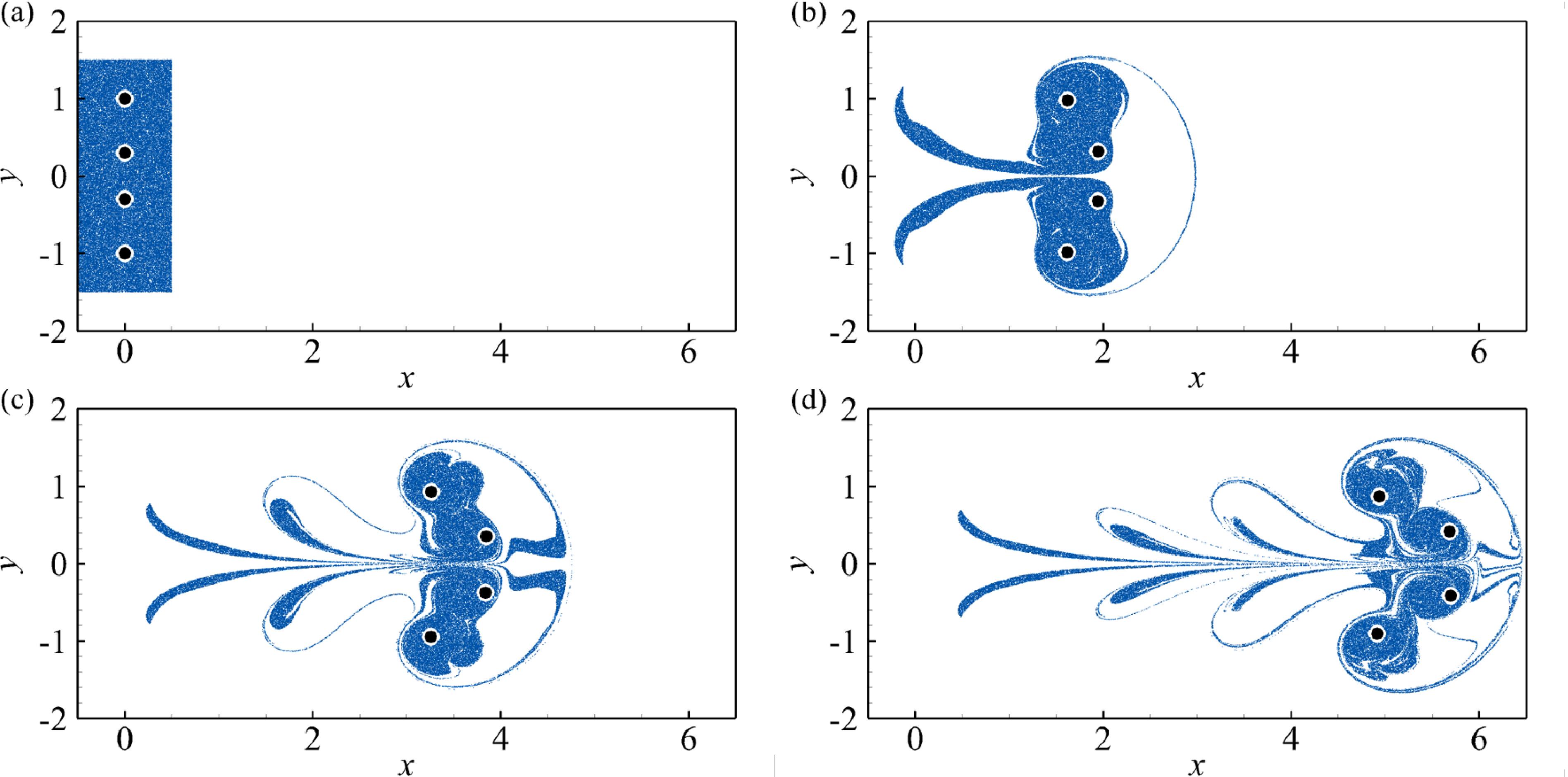}\\
  \caption{Vorticity field predicted by 4 NVM vortices under the initial condition of leapfrogging vortex rings at $t=0$, $t=11$, $t=22$, and $t=33$. Vortices are indicated by the white--black circles. For better visualization, we use 80000 tracers.}
  \label{fig:leap_frog}
\end{figure}
\begin{figure}
  \centering
  \includegraphics[width=1.0\textwidth]{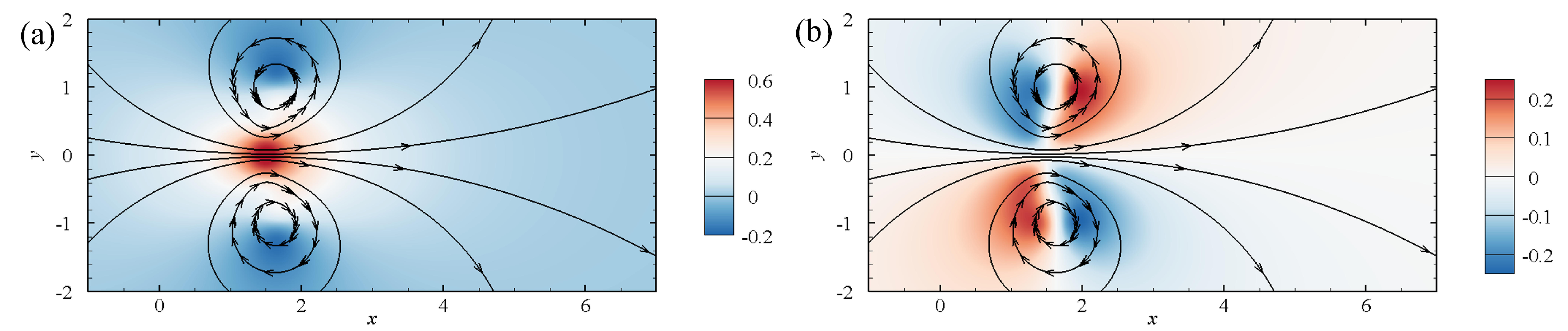}\\
  \caption{Velocity field of the leapfrog vortex at $t = 10$. The black arrow lines are the fluid streamlines. The contours represent the velocity components in (a) x-direction and (b) y-direction, respectively.}
  \label{fig:leapfrog_velocity}
\end{figure}

A classic example of interesting vortex dynamics is the
leapfrogging vortex rings, which is an axisymmetric laminar flow.
This example demonstrates our method's ability to keep the fluid field's symmetric structure. Here, we use NVM to predict the motions of 4 vortices and add 80000 randomly initialized tracers for better visualization. Since the tracers do not affect the dynamics of the underlying vorticity field, we use BS law to calculate the motions of these tracers for faster visualization. In the prediction, we initialize four vortex particles at $\bm X_1 = (0,1)$, $\bm X_2 =(0,-1)$, $\bm X_3 = (0,0.3)$, and $\bm X_4 =(0,-0.3)$, respectively, where the corresponding particle strength are $\Gamma_1=0.8$, $\Gamma_2 =-0.8$, $\Gamma_3 = 0.8$, and $\Gamma_4 = -0.8$. Although the time span of the training samples is $T_{train}=1$, we predict the evolution of vortex dynamics with a time span $T_{predict}$ of more than 30. Figure \ref{fig:leap_frog} shows the evolution of the vorticity field predicted by NVM under the initial condition of leapfrogging vortex rings at $t=0$, $t=11$, $t=22$, and $t=33$. NVM accurately captures the symmetric structure of the leapfrogging vortex rings without losing such features as time evolves. In addition, the velocity field at the $t=10$ is visualized in Figure \ref{fig:leapfrog_velocity}. Here we show the spatial distribution of the direction and magnitude of the velocity field. We can further see that the NVM maintains the intrinsic symmetry of the leapfrog vortex flow.

\begin{figure}
  \centering
  \includegraphics[width=1.0\textwidth]{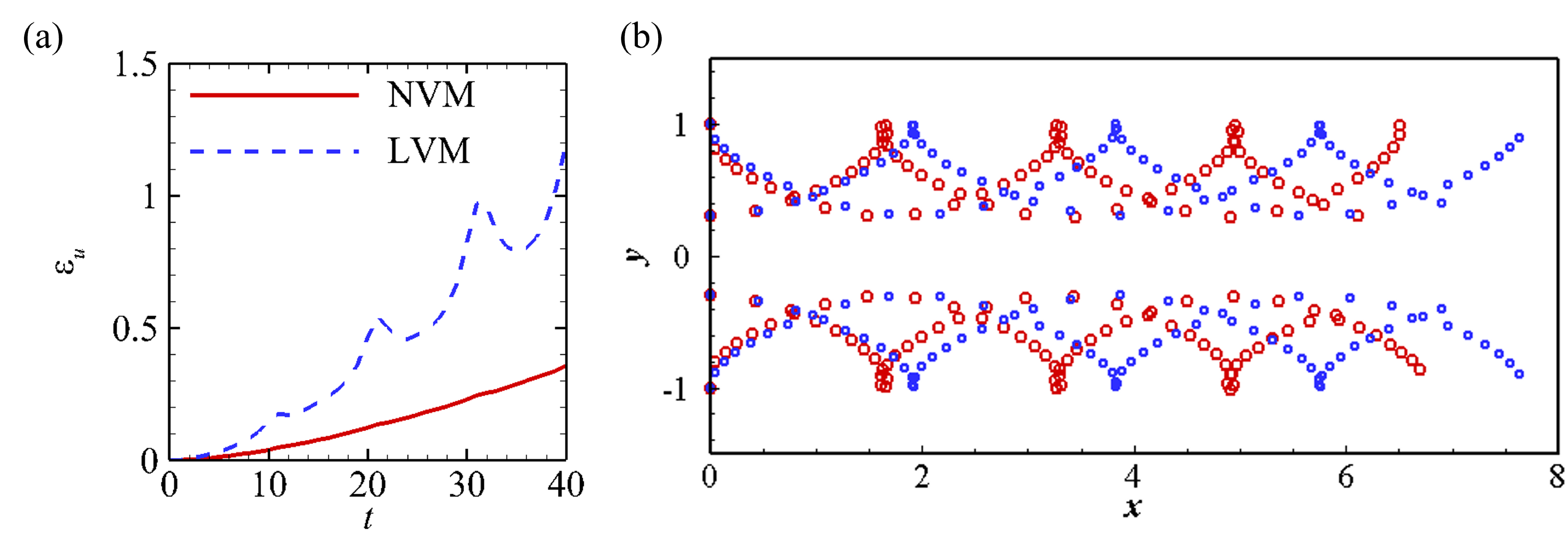}\\
  \caption{Comparison of NVM and LVM for solving leapfrog vortex flow with same number of vortex particles. (a) The relative error of velocity in flow simulation. (b) The trajectories of four vortex particles at different times from $t = 0$ to 40.}
  \label{fig:errorleap}
\end{figure}

In the vortex evolution simulated by the DNS grid method, the spatially continuous vortices are distorted into spiral structures during the interaction, and thus the mutual induction velocity of leapfrog vortices is slowed down accordingly. The extent of this evolution cannot be achieved by modeling the mathematical equations of only a few vortex particles.
In Figure \ref{fig:errorleap}, we compare NVM and LVM to solve leapfrog vortex flow. LVMs require a large number of vortex particles to accurately simulate such vortex interactions. We use neural networks to learn the intrinsic dynamics to accurately reconstruct the complex vortex dynamics with a small number of vortex particles in a data-based manner.

\subsection{Turbulent flows}

Besides simple systems like leapfrogging vortex rings, NVM is capable of predicting complicated turbulence systems.
This example's primary purpose is to illustrate our network's ability to handle more complex problems.

\begin{figure}
  \centering
  \includegraphics[width=1.0\textwidth]{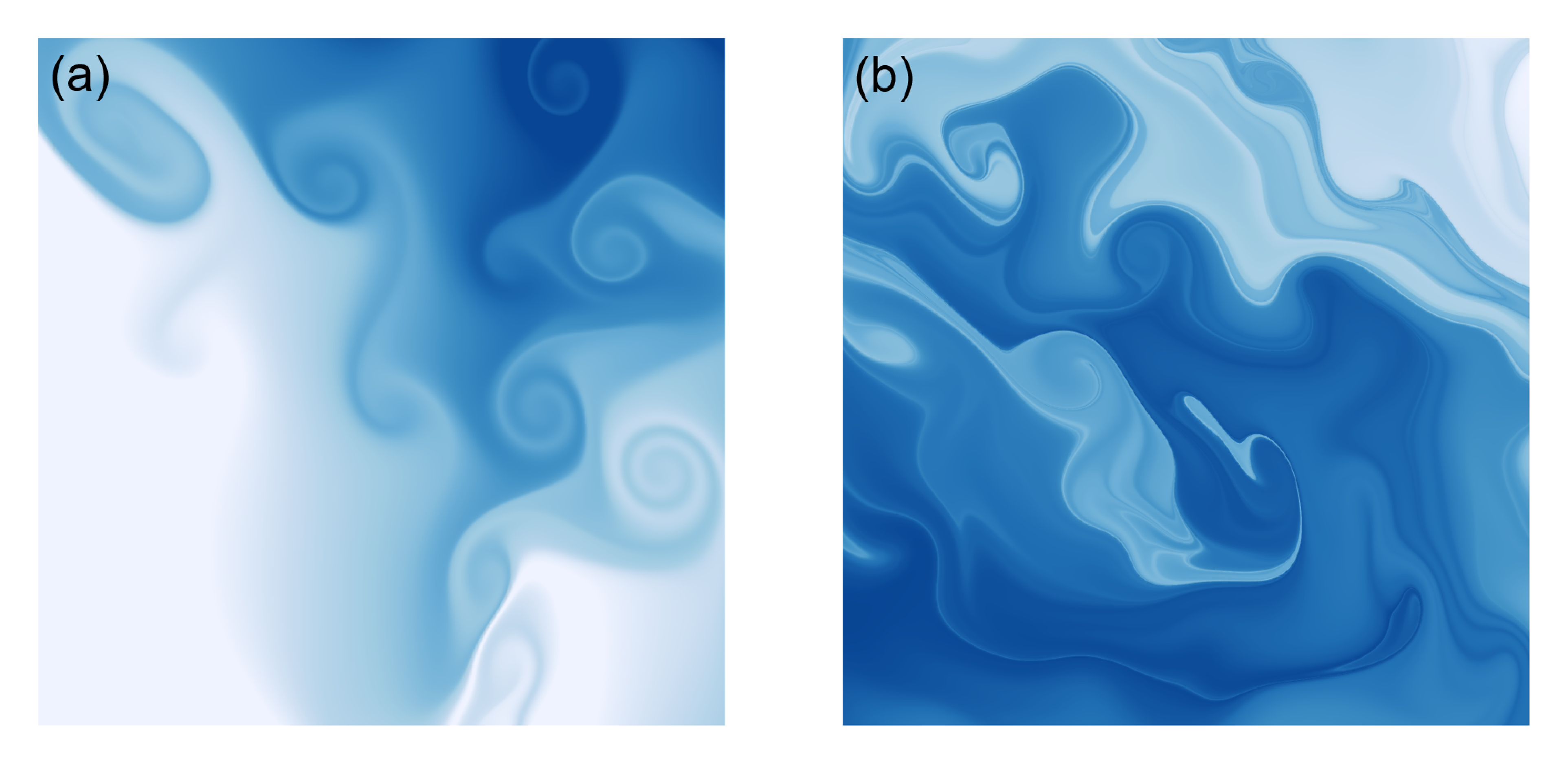}\\
  \caption{Two-dimensional Lagrangian scalar fields at $t=1$ with the initial condition $\phi = x$ and resolution $2000^2$. The evolution of the Lagrangian scalar fields is induced by (a) $O(10)$ and (b) $O(100)$ random NVM vortex particles.}
  \label{fig:turbulence}
\end{figure}

Figure \ref{fig:turbulence} depicts the two-dimensional Lagrangian scalar fields at $t=1$ with the initial condition $\phi = x$ and resolution $2000^2$.
The governing equation of the Lagrangian scalar fields is
\begin{equation}
    \frac{\partial \phi}{\partial t}+\bm u \cdot \bm \nabla \phi = 0.
    \label{eq:phi_L}
\end{equation}
The evolution of the Lagrangian scalar fields is induced by $O(10)$ and $O(100)$ NVM vortex particles at random positions $\backsim U(0,4)$ with random strengths $\backsim U(0,2)$.
We remark that the same trained model is used for both cases. There is no correlation between the positions and vortex particle strengths of the two sets of vortex particles.

Based on the particle velocity field from the NVM, a backward-particle-tracking method is applied to
solve \eqref{eq:phi_L}. Then the iso-contour of the Lagrangian field can be extracted as material structures in the evolution \cite{Yang2010a,Yang2011a,ZhaoYangChen2016a,ZhengYangChen2016,Zheng2019}. In Figure \ref{fig:turbulence} (a), the spiral structure \cite{Lundgren1982, Lundgren1993} of individual NVM vortex particles can be observed clearly due to the small number of NVM vortex particles. In Figure \ref{fig:turbulence} (b), the underlying field exhibits turbulent behaviors since it is generated with a large number of NVM vortex particles.

Generally, the high-resolution results shown in Figure \ref{fig:turbulence} can only be achieved by supercomputation using grid-based methods \cite{Yang2010a}, while NVM allows these to be generated on any laptop with GPU. 
We demonstrate that NVM is capable of generating an accurate depiction of complex turbulence systems with low computational costs.

\begin{figure}
  \centering
  \includegraphics[width=0.5\textwidth]{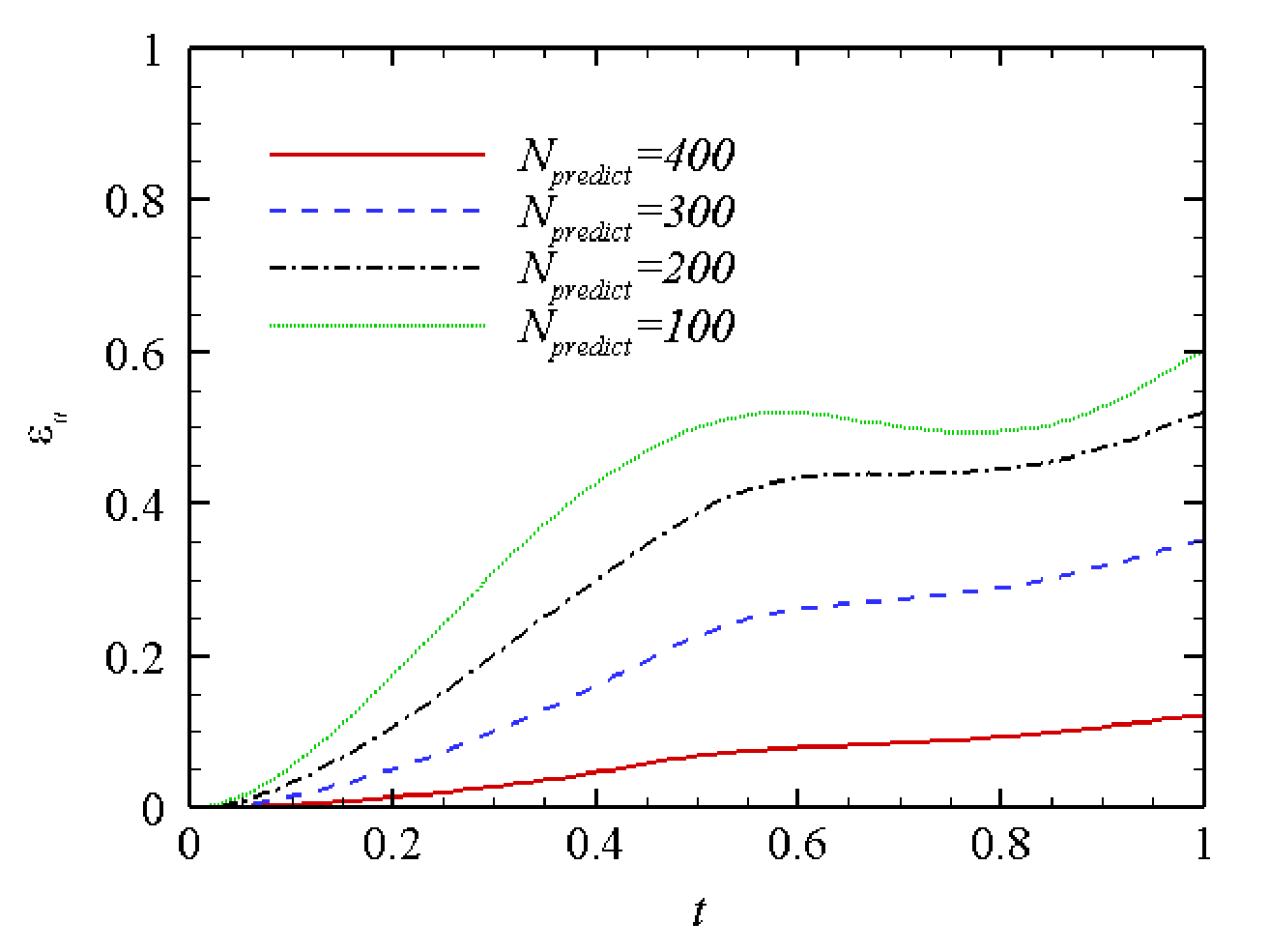}\\
  \caption{The relative errors of velocity for predicting the problem in Figure \ref{fig:turbulence}(b) with different particle sizes. Here $N_{predict}$ is the number of NVM particles used to predict the fluid evolution.}
  \label{fig:errorturbulence}
\end{figure}

We show in Figure \ref{fig:errorturbulence} the relative errors of velocity for predicting the problem in Figure \ref{fig:turbulence}(b) with different particle sizes. It can be seen that the flow evolution predicted by the NVM converges to the exact solution simulated by the DNS grid method as the number of NVM particles increases.

\subsection{NS equations with different external forces} \label{sec:threeforce}

\begin{figure}
  \centering
  \includegraphics[width=1.0\textwidth]{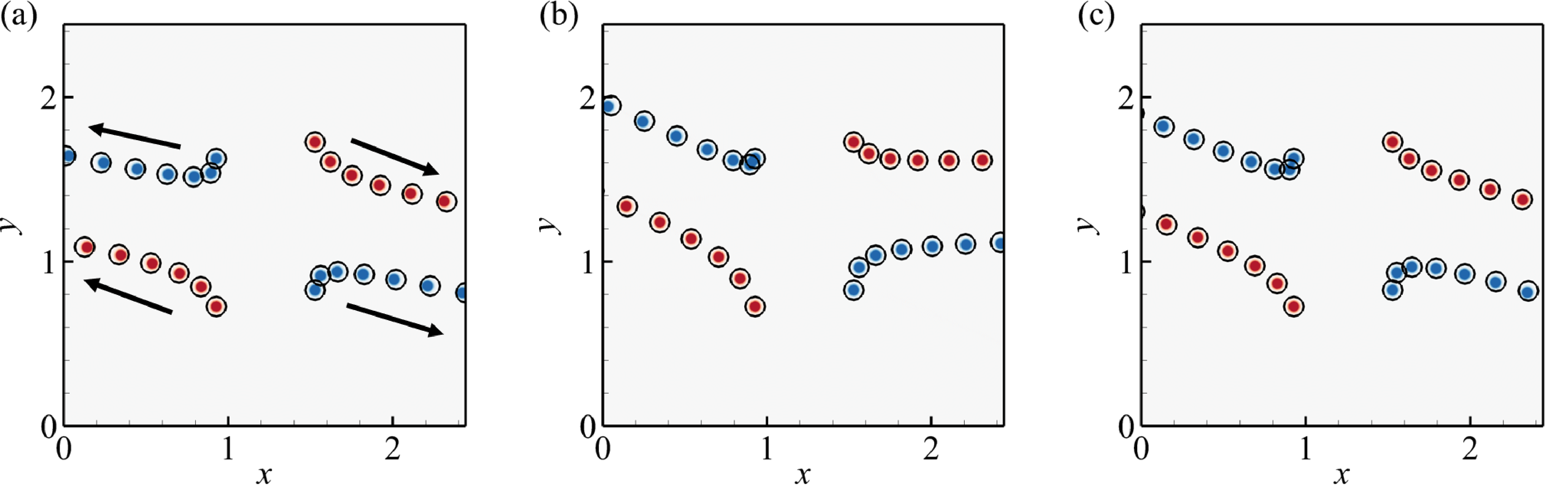}\\
  \caption{Prediction results using NVM for three cases with (a) $\bm f = \bm 0$, (b) $\bm f = (0.05\omega,0)$, and (c) $\bm f = 0.02\omega[\cos(x-x_c),-\sin(y-y_c)]$. The black arrows indicate the directions of the motions of the 2 vortices. $\omega$ represents the vorticity, and $(x_c,y_c)$ is the center of the computation domain. }
  \label{fig:predict123}
\end{figure}

In Figure \ref{fig:predict123}, we show NVM's ability to stably make accurate predictions of fluid dynamics governed by NS equations with different external forces, which are
\begin{equation}
    \begin{dcases}
    \bm f = \bm 0,\\
    \bm f = (0.05\omega,0),\\
    \bm f = 0.02\omega[\cos(x-x_c),-\sin(y-y_c)],
    \end{dcases}
\end{equation}
where $\omega$ represents the vorticity, and $(x_c,y_c)=(\pi,\pi)$ is the center of the computation domain. In the prediction, we initialize four vortex particles at $\bm X_1 = (\pi - 0.3,\pi -0.5)$, $\bm X_2 = (\pi + 0.3,\pi -0.4)$, $\bm X_3 = (\pi + 0.3,\pi +0.5)$, and $\bm X_4 = (\pi - 0.3,\pi +0.4)$, where the corresponding particle strength are $\Gamma_1=0.75$, $\Gamma_2 =-0.75$, $\Gamma_3=0.75$, and $\Gamma_4 =-0.75$. Here, we did not plot the results generated by LVM because LVM is incapable of solving the fluid dynamics with nontrivial external forces since it fails to capture the drift velocity $\bm v_i$ caused by the external force.

\begin{figure}
  \centering
  \includegraphics[width=1.0\textwidth]{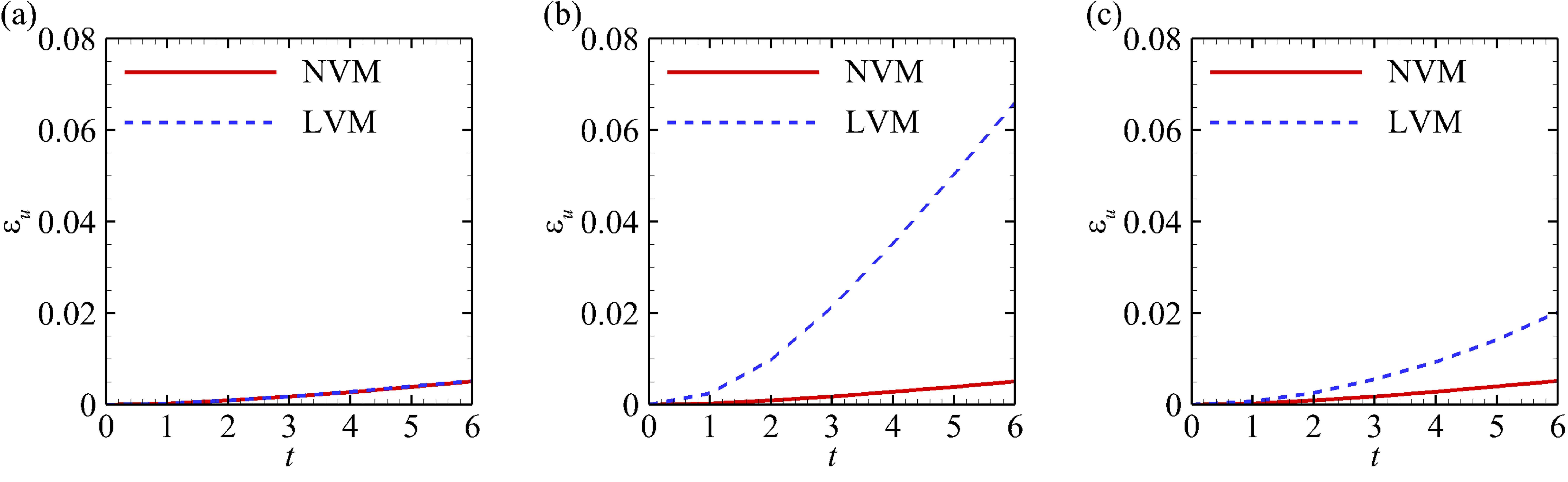}\\
  \caption{The comparison of relative error
of velocity in flow simulation using LVM and NVM for three cases with (a) $\bm f = \bm 0$, (b) $\bm f = (0.05\omega,0)$, and (c) $\bm f = 0.02\omega[\cos(x-x_c),-\sin(y-y_c)]$.}
  \label{fig:error3}
\end{figure}

We compare the relative errors of velocity in the Figure \ref{fig:error3}. It shows that for the case in Figure \ref{fig:error3}(a) without external force, NVM and LVM behave similarly, which is also related to the relatively short evolution time. Despite this, we can see that LMV shows a rapid error increase in Figures \ref{fig:error3}(b) and \ref{fig:error3}(c).

 \section{Conclusion}\label{sec:concl}
We propose a novel learning-based framework, NVM, which builds a neural-network description of the Lagrangian vortex structures and their interaction dynamics to reconstruct the high-resolution Eulerian flow field in a physically-precise manner. We demonstrate the efficacy of our method in generating highly accurate prediction results, with low computational cost, for the leapfrogging vortex rings system, the turbulence system, and the systems governed by NS equations with different external forces. Our method is the first approach to utilizing the motions of finite particles to learn infinite dimensional dynamical systems. Featured by its unique ability to generate highly accurate prediction results with low computational cost, NVM marks a significant advancement in numerical fluid simulation.

Our method has the following limitations. We train the network with a vorticity field data set rather than an experimentally measured velocity field data set. While ideally the latter is an interesting and promising study worth exploring in the future, given that more extensive data sets and complex network structures are probably required to train the network using velocity field data sets, we have not ventured into this area yet. In addition, for the particle system in our neural network, the computational cost $O(N^2)$ of the particle interactions limits larger-scale numerical simulations.

In the future, we will explore a broader range of applications of NVM. Our method only considers two-dimensional flows. In fact, natural flows have high-dimensional complexity and pose a more significant challenge to the training of detection neural networks. Additionally, for flows with stationary or moving boundaries, the vortex structures are concentrated near a thin boundary layer, and We should design neural networks that take into account the boundaries in a general network framework.
Moreover, since our detection network takes pictures as input, an interesting direction is to use high-resolution experimental flow fields as training data sets and use the trained model to predict fluid evolution.

\section*{Data Availability Statement}
The data supporting this study's findings are available from the corresponding author upon reasonable request.

\section*{Declaration of Competing Interest}
The authors declare that they have no known competing financial interests or personal relationships that could have appeared to influence the work reported in this paper.

\section*{Acknowledgements}
We acknowledge the funding support from NSF-1919647, 2106733, 2144806, 2153560, Neukom Institute CompX Faculty Grant, and Burke Research Initiation Award.

\bibliographystyle{unsrt}
\bibliography{refs}

\end{document}